\documentstyle[preprint,revtex]{aps}

\begin{document}

\begin{title}
Relativistic  three-particle dynamical equations: II. Application to \\ the
trinucleon system
\end{title}
\author{Sadhan K. Adhikari and Lauro Tomio}
\begin{instit}
Instituto de F\'\i sica Te\'orica, Universidade Estadual Paulista \\
01405 S\~{a}o Paulo, S\~{a}o Paulo, Brasil
\end{instit}

\begin{abstract}

We calculate the contribution of relativistic dynamics on the neutron-deuteron
scattering length and triton binding energy employing five sets
trinucleon potential models and four types of three-dimensional
relativistic three-body equations suggested in the preceding paper.
The relativistic correction to binding
energy may vary a lot and even change sign depending on the relativistic
formulation  employed. The deviations of these observables
from those obtained in nonrelativistic models follow the general universal
trend of deviations introduced by off- and on-shell variations of two- and
three-nucleon potentials in a nonrelativistic model calculation.
Consequently, it will be difficult to separate
unambiguously the effect of off- and on-shell variations of two- and three-
nucleon potentials on low-energy three-nucleon observables from the effect
of relativistic dynamics.
\vskip 1.2cm
PACS numbers:{25.10.+s, 21.45.+v, 24.10.Jv}
\end{abstract}


\newpage

\section{INTRODUCTION}
As the two-nucleon observables test the two-nucleon potential only on-shell
one needs to consider the few-nucleon system to test the off-shell properties
of this potential. Also, one needs to consider the few-nucleon system in order
to study the effect of the three-nucleon interaction.
There has been a great deal of experimental and theoretical activities
in the three-nucleon system  over the last three decades with the objective
of extracting informations about the two- and three-nucleon interactions. In
the recent past there has been many benchmark calculations involving
realistic two- and three-nucleon potentials.\cite{bench,bench2,ak}
Though it has been possible to fit most of the low-energy three-nucleon
observables using an appropriate
${\it ad}$ ${\it hoc}$ mixture of reasonable two- and three-nucleon
potentials, not much physics was learnt from these calculations. No
reasonable criteria for preferring one nonrelativistic
meson-theoretic\cite{meson} potential model over
another for this system has been obtained from these calculations. Though
these calculations have been successful in explaining a great deal of
experimental data, they have revealed very little new information about the
two- and three-nucleon interactions, once the potential models satisfy
some reasonable criteria, such as possessing the correct tail.\cite{ak}

The most easily and commonly studied three-nucleon observables, which are very
sensitive to variations of two- and three-nucleon interactions,  are the triton
binding energy, $B_t$, and the $S$-wave spin-doublet
neutron-deuteron scattering length, $a_{nd}$. Quite sometime ago
Phillips\cite{ak,phillips} noted that, in nonrelativistic potential model
calculations, these two observables are always correlated. Later many other
correlations have been observed in the $S$-wave spin-doublet observables.
Girard and Fuda\cite{fuda} found that the $S$-wave asymptotic
normalization parameter of triton is correlated with $B_t$ or $a_{nd}$.
A correlation has been observed between the r.m.s. radius of triton and
$B_t$.\cite{bench,bench2,ak} There has been
correlations involving the $D$-state observables of the three-nucleon
system.\cite{bench,bench2,ak}

If two three-nucleon nonrelativistic dynamical models yield the same value for
$B_t$ or $a_{nd}$ they should yield identical results for many other
correlated three-nucleon observables.\cite{bench,bench2,ak,phillips}
These observables of the three-nucleon system, which exhibit the correlated
behavior, are usually most sensitive to the variations of the three-nucleon
 potential models.
The low-energy correlations make it simple to classify the
results of theoretical calculations, while at the same time make the
extraction of physically meaningful information that much harder.\cite{ak}

The importance of relativistic effects in the three-nucleon calculations
has never been overemphasized. Both the bound-state and low-energy
scattering calculations
involve large momentum components which demand a relativistic
dynamical treatment of the problem. Relativistic dynamical calculations in
the three-nucleon problem have been mainly restricted to the study of
the three-nucleon bound state problem\cite{old,rupp1,rupp2,garci1,mach}
with one exception where relativistic effect on the neutron-deuteron
scattering length has been studied\cite{garci2}.  However,
the objective of all these studies has been  the same. The authors have been
mainly concerned in explaining the missing gap between the predictions of
a nonrelativistic potential model for the three-nucleon system and experiment
by incorporating some kind of relativistic dynamics. Both the four-dimensional
Bethe-Salpeter-Faddeev equation\cite{rupp1,rupp2} in some approximate form
and several types of three-dimensional reductions of this equation have
been employed for this purpose.\cite{old,rupp1,rupp2,garci1,mach,garci2}

Though the magnitude of relativistic corrections to $B_t$ and $a_{nd}$,
as emphasized in previous studies, is interesting, in our opinion it is
most relevant to see if meaningful physics could be extracted from the
relativistic treatment of the three-nucleon system. The nonrelativistic
potential model calculations of the three-nucleon system involving
meson-theoretic nucleon-nucleon potentials\cite{meson} did not allow us
to extract meaningful informations about the two- and three-nucleon
interactions because of the correlated behavior of the observables directly
sensitive to these interactions.\cite{ak} The  question to ask at this stage is
whether the relativistic treatment of the three-nucleon problem  is expected
to change the scenario.

It is still unclear on how to progress from QCD to practical collision
integral equations for hadronic and nuclear processes. Nevertheless, often
for hadronic systems a Bethe-Salpeter (BS)
 type equation is postulated using some type of
 meson-baryon field theory with phenomenology, that presumably have a
 wider range of validity than nonrelativistic equations of the
 Lippmann-Schwinger (LS) type.  Usually, the ladder approximation to the BS
 equation and its subsequent reduction to three-dimensional form
\cite{bls,aaron,ahmad,our}   have permitted  numerical calculations.
 It is
reasonable to require that all the approximate versions of the BS equation
satisfy conditions of time-reversal symmetry, unitarity, and relativistic
covariance. One of the approximate versions considered so far\cite{ahmad}
and frequently used in numerical calculations\cite{old,rupp1,rupp2,mach} in
an approximate form do
not even satisfy conditions of time-reversal symmetry.
However, at present time, in spite of these defects, one of the practical and
feasible ways for performing a relativistically covariant three-nucleon
calculation is through some of these approximate three-dimensional
equations and we use them for studying the relativistic effect to the
three-nucleon problem.  At this point it should be noted that the
solution of the approximate BS equation in ladder form is not necessarily
 a superior way of dealing with the relativistic effect.\cite{mach,gross}

In order to verify if new informations about two- and three-particle
interactions could be obtained via a relativistic dynamical three-nucleon
calculation, we have performed three-nucleon calculation for $B_t$ and
$a_{nd}$ using several separable potential models and four approximate
versions of three-dimensional relativistic three-particle equations suggested
in the preceding paper.\cite{our}
The spin variables are treated nonrelativistically. We do not pretend to
claim that the separable potential model presents a realistic description
of the three-nucleon system. However, the numerical calculation is
simplified by an order of magnitude in this model, and  this model has
been  used successfully in understanding the essential features of the
nonrelativistic three-nucleon problem. Here we employ the  relativistic
version of the three-particle separable potential model with a hope to see
if new physics could be extracted from a study
of the low-energy observables of the three-nucleon system.

   We employ Yamaguchi and Tabakin-type\cite{yt}
   nucleon-nucleon $^3S_1$ and $^1S_0$
potentials in the present calculation. Tabakin-type nucleon-nucleon potentials
yield nucleon-nucleon phase shifts in better agreement with experiment, which
change sign at higher energies, compared to the Yamaguchi potential. If
Tabakin-type potential is used in both $^3S_1$ and $^1S_0$ spin channels, it
leads to an unrealistic triton ground state of several hundred MeV's.\cite{bsc}
The use of the Tabakin potential in one of the nucleon-nucleon
spin channels and Yamaguchi in the other, as has been done in the present
calculation,  does not lead to a collapsed triton and lead to trinucleon
observables in better agreement with experiment and realistic calculations.

We derive certain general theoretical inequalities among  the different
triton binding energies  obtained using nonrelativistic and various
relativistic  dynamical formulations. These inequalities are verified in
actual numerical calculations and are expected to be valid in general for
other potential models. All the
relativistic models satisfy conditions of relativistic covariance and
unitarity. As there is no obvious theoretical reason for prefering one of
the relativistic
formulations over another, in view of these inequalities it is not to the
point to talk about the absolute value of the relativistic corrections to
$B_t$ or $a_{nd}$; one could have corrections of different magnitudes and
signs.

We present the nonrelativistic and  relativistic three-nucleon models, which
we use in numerical calculations, in Sec. II. Numerical results are
presented in Sec. III and finally, a summary of our findings are given in
Sec. IV.

\section{Dynamical Models}
As we shall only be considering the three-nucleon system, it is convenient to
consider three equal-mass particles of mass $m$, where $m$ is the nucleon mass.
In our calculation we use $\hbar c =197.33$ MeV fm, and $m =938.97$ MeV.

The nonrelativistic two-nucleon dynamics for a central $S$ wave potential
is governed by the following
partial-wave Lippmann-Schwinger (LS) equation
\begin{equation}
t(q',q,k^2)=V(q',q)+4{\pi}\int_0^\infty p^2 dp V(q',p)\frac{1}
{k^2-p^2+i0} t(p,q,k^2),
\label{A} \end{equation}
where $V(q',q)$ is the usual momentum space potential.
The relativistic two-nucleon dynamics
for the same potential is taken to be governed by the following
partial-wave Blankenbecler-Sugar (BlS)  equation\cite{bls}
\begin{equation}
t(q',q,k^2)=V(q',q)+4{\pi}\int_0^\infty p^2 dp  \frac{m}{\omega_p}
V(q',p)\frac{1} {k^2-p^2+i0}t(p,q,k^2),
\label{B} \end{equation}
where  $\omega_p =(m^2+p^2)^{1/2}$. Equation (\ref{B})
satisfies the conditions of relativistic unitarity and covariance. However,
these conditions are not enough to specify the relativistic dynamics properly.
Actually, there are a host of such equations.\cite{ak,our}
In our study, however, at the two-nucleon level we shall only
consider the dynamics given by BlS Eq. (\ref{B}).

We shall consider only separable forms for two-nucleon potentials. There
is a convenient way of defining phase equivalent
nucleon-nucleon potentials using relativistic and nonrelativistic equations.

We take the relativistic nucleon-nucleon
potential of  the following form
\begin{equation}
  [V_n(q',q)]_{rel} = -\lambda _n [v_n(q')]_{rel} [v_n(q)]_{rel},
\label{1} \end{equation}
where $n=0$ (1) represents the spin triplet (singlet) state, and the
subscript  $rel$ ($nr$) denotes relativistic (nonrelativistic). Several
analytic form factors have been used for the form factor
$[v_n(q)]_{rel} [\equiv N_n g_n(q)]$,
where $N_0$ ($N_1$) is the normalization for the momentum space spin
triplet deuteron (singlet virtual-state)
wave function $\phi(q)= N_0 g_0(q) (\alpha_0 ^2+q^2)^{-1}$. Here, $\alpha_0^2$
is the triplet
deuteron binding energy in fm$^{-2}$; similarly, $\alpha_1^2$ is the
singlet virtual state energy.

  The relativistic $t$ matrix in this case at the square of the center of
mass (c.m.)  energy $s=4(m^2+k^2)$ is given by
\begin{equation}
[t_n (q',q,k^2)]_{rel} = [v_n(q')]_{rel} [\tau_n^{-1}(k^2)]_{rel}
 [v_n(q)]_{rel},
\label{3} \end{equation}
where
\begin{equation}
[\tau_n (k^2)]_{rel} = -\frac{1}{\lambda _n} - 4{\pi}\int^{\infty}_0
q^2 dq \left( \frac{m}{\omega_q}\right) \frac {[v_n(q)]^2_{rel} }{k^2-q^2+i0}.
\label{4} \end{equation}
 We  generate a nonrelativistic  two-nucleon
$t$ matrix, phase-equivalent to its relativistic version by
the following transformation for the form-factors
\begin{equation}
[v_n(q)]_{nr} = (\sqrt {m/\omega_q}) [v_n(q)]_{rel},
\label{13} \end{equation}
 so that
 \begin{equation}
 [t_n(q',q,s)]_{nr} = [v_n(q')]_{nr}[\tau_n^{-1}(k^2)]_{rel}[v_n(q)]_{nr},
 \label{14} \end{equation}
 The functional form of
  $[\tau]_{rel}$ of Eq. (\ref{14})
 is exactly identical to its relativistic counterpart (\ref{4}).
\vskip .5cm

The above recipe generates
phase-equivalent two-nucleon potentials to be used in nonrelativistic and
relativistic  three-nucleon problem.  The nonrelativistic and
relativistic versions lead to the same deuteron binding $\alpha _0^2$
in units  of fm$^{-2}$.
However,  one uses a distinct relation for transforming this energy to MeV
in relativistic and nonrelativistic versions.
 Consequently, the relativistic and nonrelativistic deuteron
bindings are slightly different.

The nonrelativistic Faddeev equations for the three-nucleon system
is given by\cite{ak}
\begin{eqnarray}
\Xi_{n,n'}(p,p',E)  & = & Z_{n,n'}(p,p',E) +\sum_l \int_0^{\infty}
q^2 dq Z_{n,l}(p,q,E) \left[-\frac{3}{2\pi}\tau_l^{-1}(mE-3q^2/4)\right]
\nonumber \\
& \times &  \Xi_{l,n'}(q,p',E),
\label{5} \end{eqnarray}
with
\begin{equation}
Z_{n,n'}(p,q,E) = \frac{8\pi^2}{3} J_{n,n'}\int_{-1}^{1} dx
[v_n({\cal P})]_{nr} G_{nr}(\vec p,\vec q,E) [v_{n'}({\cal Q})]_{nr},
\label{6} \end{equation}
where $G_{nr}$ is the three-particle nonrelativistic propagator given by,
\begin{equation}
G_{nr}(\vec p,\vec q,E) =(p^2+q^2+pqx -mE -i0)^{-1},
\label{7} \end{equation}
with
\begin{equation}
  {\cal P}^2 = p^2/4 +q^2 +pqx,
  \label{8} \end{equation}
  and
\begin{equation}
 {\cal Q}^2 = q^2/4 +p^2 +pqx.
 \label{9} \end{equation}
Here $J'$s are the spin-isospin recoupling factors given by $J_{00}
=J_{11} = 1/4$, and $J_{01}=J_{10}=-3/4$ for the spin doublet system. The
scattering length in this case is given by $a_{nd}
=-\Xi_{0,0}(0,0,mE=-\alpha _0^2)$.

The three-dimensional relativistic generalization of these Faddeev equations
has a form similar to Eq.(\ref{5}) and is given by\cite{bls,aaron,our}
\begin{eqnarray}
\Xi_{n,n'}(p,p',s) & = & Z_{n,n'}(p,p',s) +\sum_l \int_0^{\infty}
q^2 dq \frac {m} {\omega _q} Z_{n,l}(p,q,s)
\left[ -\frac{3}{2\pi}\tau_l^{-1}[(s-3m^2-2\omega_q \sqrt s)/4]\right]
\nonumber \\
& \times  & \Xi_{l,n'}(q,p',s),
\label{20} \end{eqnarray}
and Eq. (\ref{6})
but  with the relative momentum squares given by Eqs. (\ref{8}) and (\ref{9})
now changed to the following relativistic forms:
\begin{equation}
  {\cal P}^2 = (\omega _q+\omega_{pq})^2/4-p^2/4 -m^2,
  \label{21} \end{equation}
 \begin{equation}
  {\cal Q}^2 = (\omega _p+\omega_{pq})^2/4-q^2/4 -m^2.
\label{22} \end{equation}
Here we use notations $\omega_p =(m^2+p^2)^{1/2}$, $\omega_{pq}
 =[m^2+(\vec p +\vec q)^2]^{1/2}$, etc. It should be noted that in the
nonrelativistic limit Eq. (\ref{20}) reduces to Eq. (\ref{5}).

 Finally, for the relativistic three-particle propagator $G$ we use the
 following functions:\cite{aaron,our}
 \begin{equation}
 G_A(\vec p,\vec q,s) = \frac {2(\omega_p+\omega_q+\omega_{pq})} {\omega_{pq}
 [(\omega_p+\omega_q+\omega_{pq})^2 - s-i0]};
 \label{23} \end{equation}
 \begin{equation}
 G_B(\vec p,\vec q,s) = \frac {2(\omega_p+\omega_q)} {\omega_{pq}
 [(\omega_p+\omega_q)^2 - (\sqrt s -\omega_{pq})^2-i0]};
 \label{24} \end{equation}
 \begin{equation}
 G_C(\vec p,\vec q,s) = \frac {1} {\omega_{pq}
 [\omega_p+\omega_q+\omega_{pq} - \sqrt s-i0]};
 \label{25} \end{equation}
 \begin{equation}
 G_D(\vec p,\vec q,s) = \frac {2 (\omega_q+\omega_{pq})} {\omega_{pq}
 [(\omega_q+\omega_{pq})^2 - (\sqrt s - \omega_p)^2-i0]}.
 \label{26} \end{equation}
 In Eqs. (\ref{23})
- (\ref{26}) the parameter $s$ is the square of the total
c.m. energy of the three-particle system. All these propagators satisfy
conditions of relativistic unitarity, governed by that part of the denominator
in these propagators which corresponds to the pole for three-particle
propagation in the intermediate state, e.g., at
$\sqrt s = \omega_p+\omega_q+\omega_{pq}$.  The condition of relativistic
unitarity in these propagators is manifested in having the same residue at
this pole.

All these equations satisfy two-particle unitarity via the use of the
BlS equation. Equation (\ref{23}) was implicit in the work of BlS but was
explicitly advocated by Aaron, Amado, and Young\cite{aaron}
and obeys time-reversal symmetry, e.g. $G(\vec p,\vec q,s) =
G(\vec q,\vec p,s)$, and both two- and three-particle unitarity.
Equations (\ref{24}) and (\ref{25}) also have these virtues of Eq. (\ref{23}).
The propagators $G_B$ and $G_D$ were  suggested recently
in Ref. \cite{our}, $G_C$ was suggested long ago.\cite{ahmad} It has been
shown\cite{new} that the propagator $G_D$ follows from a suggestion
 by Ahmadzadeh and Tjon.\cite{ahmad} But in numerical applications of  this
propagator to the three-nucleon problem unnecessary nonrelativistic
approximations have been used which violate conditions of unitarity.\cite{new}
The form (\ref{26}) obeys three-particle unitarity, but  violates
time-reversal symmetry.

We have used all these forms, (\ref{23}) - (\ref{26}), in  numerical
calculation.  As the propagators $G(\vec p,\vec q,s)$'s directly enter the Born
term
of the scattering equation, useful inequalities for the three-particle  binding
energies could be obtained with these propagators, which are later verified
in numerical calculations.  For example, from Eqs. (\ref{23}) - (\ref{26})
we have
\begin{equation}
G_A(\vec p,\vec q,s) = G_C(\vec p,\vec q,s) \frac
{2(\omega_p+\omega_q+\omega_{pq})}
 {\omega_p+\omega_q+\omega_{pq}+\sqrt s},
 \label{27} \end{equation}
\begin{equation}
G_B(\vec p,\vec q,s) = G_C(\vec p,\vec q,s) \frac {2(\omega_p+\omega_q)}
 {\omega_p+\omega_q-\omega_{pq}+\sqrt s},
 \label{28} \end{equation}
\begin{equation}
G_D(\vec p,\vec q,s) = G_C(\vec p,\vec q,s) \frac {2(\omega_{pq}+\omega_q)}
 {\omega_{pq}+\omega_q-\omega_p+\sqrt s},
 \label{29} \end{equation}
The propagators $G$'s are directly proportional to the potentials in the
three-nucleon system. It should be noted
 that both for  three-nucleon bound-state and
threshold scattering problems $s \simeq 3m$ and  the variables $p$
and $q$ in Eqs. (\ref{27}) - (\ref{29}) run from 0 to $\infty$.
Consequently, $\omega _p$, $\omega _q$, and $\omega _{pq}$ are  larger
than $m$ in this domain, and the factors multiplying $G_C(\vec p,\vec q,s)$ in
Eqs. (\ref{27}) - (\ref{29}) are larger than one. So the propagators and the
potentials in models $A$, $B$, and $D$ are stronger
than that in the model $C$, impling $(B_t)_A > (B_t)_C$,
$(B_t)_B > (B_t)_C$, and $(B_t)_D > (B_t)_C$. From Eqs. (\ref{27}) and
(\ref{28}) one can
see that  the model potential $B$ is stronger than the model
potential $A$. Similarly, one can show that the model potential $D$ is
stronger than the model potential $A$.   Consequently,
$(B_t)_B > (B_t)_A > (B_t)_C$, and  $(B_t)_D > (B_t)_A > (B_t)_C$. These are
some useful inequalities. No such inequality could be established
 between models $B$ and $D$.

Hence we have the following useful inequalities
\begin{equation}
(i) (B_t)_B > (B_t)_A > (B_t)_C,\\
(ii)(B_t)_D > (B_t)_A > (B_t)_C,\\
\label{42} \end{equation}
which will be verified in the numerical calculation in the following section.

\section{numerical results}

For  two-nucleon separable potentials in  spin-triplet and spin-singlet
channels we take the following Yamaguchi and Tabakin form-factors,\cite{yt}
recently used by Rupp and Tjon:\cite{rupp2}
\begin{equation}
g_Y(q)=\frac {1} {q^2+\beta ^2},
\label{30} \end{equation}
\begin{equation}
g_T(q)=\frac{q^2+\nu ^2}{q^2+\gamma ^2} \times
\frac{q_c^2-q^2}{(q^2+\beta ^2)^{\kappa}}, \kappa=1.5, 2.
\label{31} \end{equation}
The Yamaguchi potential will be referred to as Y, and the Tabakin potential
with $\kappa = 1.5, 2$ will be referred to as T-1.5 and T-2, respectively.
The constants of these potentials for the triplet and the singlet
channels are slightly different from those of Rupp and Tjon and are given
in Table I.
These potentials fit the two-nucleon phase-shifts equally well as
in the work of  Rupp and Tjon. Potential (\ref{31}) provides a change of sign
of nucleon-nucleon phase shifts at higher energies in agreement with
experiment.

We calculated the triton binding, $B_t$, and the neutron-deuteron
scattering length, $a_{nd}$, in the nonrelativistic case as well as with
each of the four versions of  relativistic formulations $A-D$. Propagator $A$
has been used before in numerical calculations of the three-nucleon problem.
\cite{old,rupp1,rupp2,garci1,garci2} Propagator $B$ has been suggested only
recently\cite{our} and has  never been  used before. Propagator $C$ is the
simplest and has been known for a long time,\cite{ahmad} but to the best of our
knowledge has not been used in the three-nucleon problem.

Our results are
exhibited in Table II. From Table II it is clear that the relativistic
corrections to $B_t$ and $a_{nd}$ for various models may vary a lot,
even the sign of the relativistic correction may change in agreement with
inequality (\ref{42}). All the relativistic models increase the triton
binding energy, $B_t$,  in relation to the nonrelativistic case,
except model $C$ which reduces the binding.  The magnitude of the
relativistic correction to $B_t$ varies from 0.1 MeV to 0.7 MeV in
different situations. The magnitude
and even its sign changes when one changes the relativistic models. In view
of this, and related flexibilities of the various relativistic models,
 it may not be quite meaningful to  talk about  the
 magnitude of relativistic effect with a view to reduce the discrepancy
 between experiment and nonrelativistic theoretical model calculation.
 The theoretical inequalities (\ref{42}),
 however, hold true in all situations. In addition we observed in numerical
calculations the following general inequality
\begin{equation}
 (B_t)_D, (B_t)_B > (B_t)_A > (B_t)_{nr} > (B_t)_C.
\end{equation}

Our principal finding is exhibited in
 Fig. 1 where we  plot $B_t$ versus $a_{nd}$ for
the present nonrelativistic and relativistic model calculations, as well as
for many other nonrelativistic calculations taken from the literature.
\cite{opt,ad,data}
The relativistic calculations differ in employing different relativistic
dynamics and nucleon-nucleon potentials, the nonrelativistic
calculations differ in variations of two-nucleon potential off-shell and/or
three-nucleon potential. The trend of the relativistic
calculations is identical to that of the nonrelativistic calculations.
Hence, the effect of including relativistic dynamics in the three-nucleon
problem can not be distinguished from the effect of varying the two- and
three-nucleon potentials in  nonrelativistic calculations. Consequently,
 a relativistic treatment of these low-energy observables
may not enhance our knowledge of the underlying interactions, or dynamics.
In other words, in the low-energy three-nucleon observables, the hope
of separating the effect of on- and off-shell variations of the two-
and three-nucleon potentials from the relativistic effect, in a model
independent fashion, is remote.

Certain meson theoretic (two- and three-nucleon)
potentials\cite{meson} when used in a certain
relativistic formalism may reproduce low-energy three-nucleon observables.
But this should not be considered as a mark of superiority of this model
over others. An appropriate mixture of on- and off-shell variations of
the potentials and a relativistic formulation may reproduce certain
experimental results, but not much physics could be learned from such
studies. This happens because of the existence of a shape independent
approximation to many of the low-energy observables, such as binding
energy and scattering length, of the three-nucleon system, as in the
two-nucleon system.\cite{ak,opt}
These observables are insensitive  not only to the
variations of the shape of the potential, but also to inclusion of certain
relativistic dynamics, provided that the triton binding is reproduced.

\section{summary}
We have calculated the contribution of  relativistic effect on
the neutron-deuteron scattering length and the triton binding energy
 employing several separable nucleon-nucleon potentials and
three-particle relativistic equations. We have used combinations of Yamaguchi
and Tabakin type potentials for the singlet and triplet nucleon-nucleon
channels and four types of relativistic three-particle
scattering equations. To the best of our knowledge,
 of these equations only those by Aaron, Amado, and
Young (model $A$) has been used before in this context. Model $D$ can be
 derived from a three-particle propagator derived by Ahmadzadeh and Tjon, but
has not been used in numerical calculation before in this form. Previous
numerical calculations with this propagator used unnecessary nonrelativistic
approximations which violate conditions of relativistic covariance and
unitarity, as has been pointed out recently.\cite{new}
Models  $B$ and $C$ have not
been used in numerical calculations before. All these models satisfy
constraints of relativistic unitarity and covariance. However,  these
conditions are not enough to determine the dynamics. There still
remains a lot of flexibility which results in very different relativistic
corrections to the three-nucleon problem.   The relativistic correction could
be both positive and negative depending on the model chosen. The magnitude of
relativistic correction to triton energy varies from 0.1 to  0.7 MeV
(see, Table II),
depending on the relativistic dynamics and nucleon-nucleon potential model
employed.   We have derived certain inequalities for the binding energies of
different models which are verified in numerical calculations.

In addition to studying the relativistic corrections to $B_t$ and $a_{nd}$ we
also studied the correlations among these two observables. They exhibit a
correlated behavior in different nonrelativistic model  calculations
employing different on-shell equivalent nucleon-nucleon potentials. The
inclusion of a three-nucleon potential also does not change the  situation.
The present study of relativistic effect  is also in accordance with  this
correlation. Hence the inclusion of relativistic dynamics and three-nucleon
potential and off-shell variation of the nucleon-nucleon potential lead to
similar correlated behavior of $B_t$ and $a_{nd}$. Consequently, it will be
difficult to separate the effect of relativistic correction from the effect
of a variation
of the nucleon-nucleon potential off-shell from a study of these observables.
This confirms the existence of a shape-independent approximation to these
observables even after inclusion of the relativistic effect.\cite{opt}

Of course, there are  other observables for the three-nucleon system, which
should be directly sensitive to relativistic effect, such as the charge
form factors. Because of the presence of the possible large effect of
meson-exchange currents and of the non-nucleonic components in the nucleus,
such observables are not easily tractable, and it has so far been difficult
to draw model independent conclusion from studies of these
observables.\cite{bench,bench2}

We are aware that there is an inherent flexibility
 in deciding on the relativistic
dynamics, in treating the spin variables relativistically, and in deciding the
correct form of two- and three-nucleon potentials. We are far from exhausting
all possibilities. But the tendency of existing the shape-independent
approximation is so strong that we do not believe our conclusions to be
so peculiar as to be of no general validity. Hence a relativistic framework
may reduce the still existing discrepancy between theory and experiment,
but this may not enhance our knowledge of the three-nucleon system.

We thank Dr Tobias Frederico for critical comments and encouragements
throughout  this work.
The work is supported in part by the Conselho Nacional de
Desenvolvimento - Cient\'\i fico e Tecnol\'ogico (CNPq) of Brasil.

\newpage

\begin{table}
\centering
\begin{tabular} {|c|c|c|c|c|} \hline
{$$} & {$$} & {$Yamaguchi$} & {$Tabakin-1.5$} & {$Tabakin-2$}
\\   \hline

     & $\lambda N^2$     &   0.4012 (fm$^{-3}$)  & 1.7316 (fm$^{-1}$) & 256.20
     (fm$^{-3}$)       \\

&   $\beta$ (fm$^{-1}$)  &   1.4117 & 4.0335  & 5.1435 \\
$^3S_1$  &$\nu$ (fm$^{-1}$)  &      & 0.8067  & 0.8400   \\
   &   $\gamma$ (fm$^{-1}$)   &     & 0.7324 & 0.7534  \\
    &   $q_c$ (fm$^{-1}$)   &     &  2.1205 & 2.1205    \\
    &           &       &       &      \\
  & $\lambda N^2$ & 0.1487 (fm$^{-3}$) & 0.9455 (fm$^{-1}$)   & 216.01
(fm$^{-3}$)
    \\
 & $\beta$ (fm$^{-1}$)& 1.1560 & 4.057 & 5.074  \\
   $^1S_0$ &$\nu$ (fm$^{-1}$)  &  & 1.1643  & 1.1415   \\
     & $\gamma$ (fm$^{-1}$) &   & 0.9237 & 0.9065  \\
     & $q_c$ (fm$^{-1}$)  &       & 1.6966     & 1.6966    \\
\hline
\end{tabular}

\vspace{0.5cm}
\caption{Yamaguchi and Tabakin potential parameters $\lambda N^2, \beta,
\nu, \gamma, q_c$, etc. The quantity $\lambda N^2$ is the usual strength of
the separable potential, where $N$ is the normalization of the two-nucleon
state.  There parameters are fitted for the $^3S_1$ state to
$a=5.424$ fm,  $\alpha_0$=0.23161 fm$^{-1}$, and for the $^1S_0$ state to
$a=-23.748$ fm, and $\alpha_1$=0.03992 fm$^{-1}$.}
\end{table}

\begin{table}
\centering
\begin{tabular} {|c|c|c|c|c|c|c|} \hline
{$Potential$} & {$$} & {$nr$} & $A$ & {$B$} & {$C$} & $D$
\\   \hline

  YY   & $B_t$      &   10.65  & 10.73 & 10.92 & 10.39 &10.91       \\
 YY  & $a_{nd}$   &   -0.77 & -0.83  & -0.94 & -0.61 & -0.94  \\
  & & & & & & \\
  T-2Y  & $B_t$   &  8.06    & 8.14  & 8.34 & 7.91 & 8.30 \\
 T-2Y  &   $a_{nd}$    &  0.94   & 0.87 & 0.73 & 1.04 & 0.75 \\
  & & & & & & \\
  YT-2   &   $B_t$   &  7.69   &  7.87 & 8.30 & 7.52 & 8.19    \\
 YT-2   & $a_{nd}$  &  1.30     & 1.15      & 0.82 & 1.42 & 0.90     \\
  & & & & & & \\
  T-1.5Y   &   $B_t$   &  7.99   &  8.09 & 8.35 & 7.85 & 8.27    \\
 T-1.5Y   & $a_{nd}$  &  0.98     & 0.91      & 0.71 & 1.08 & 0.77     \\
  & & & & & & \\
  YT-1.5   &   $B_t$   &  7.59   &  7.78 & 8.38 & 7.44 & 8.17    \\
 YT-1.5   & $a_{nd}$  &  1.39     & 1.22      & 0.75 & 1.49 & 0.93     \\
\hline
\end{tabular}
\vspace{0.5cm}
\caption{Triton binding energy $B_t$ (MeV) and neutron-deuteron scattering
length
$a_{nd}$ (fm) for different nucleon-nucleon potential models (Yamaguchi,
Tabakin-1.5, and Tabakin-2) and relativistic ($A, B, C, D$) and
nonrelativistic ($nr$) dynamics. The three-nucleon potential model XY has a
triplet X and singlet Y nucleon-nucleon potential, where each of X and Y
could be Y, T-1.5, and T-2 of Eqs. (\ref{30}) and (\ref{31}).
For example, YT-2 denotes a triplet Yamaguchi and singlet
Tabakin-2 potential.}
\end{table}
\vskip 1cm
{\bf Figure Caption}\\

1. The $B_t$ versus $a_{nd}$ plot for various trinucleon models: the present
relativistic
models ($\diamond$), the present  nonrelativistic models ($+$),
and other nonrelativistic models
taken from the literature ($\times$).

\end{document}